\documentclass[12pt]{article}
\usepackage[top=1in, left = 1in, bottom = 1in, right = 1in, paperheight = 11in, paperwidth = 8.5in]{geometry}

\usepackage[utf8]{inputenc}
\usepackage{adjustbox}
\usepackage{sectsty}
\usepackage[hyphens]{url}
\usepackage[breaklinks]{hyperref}
\usepackage{amsmath, amssymb, amsthm, amsbsy, amsfonts, cases}
\usepackage{graphicx, comment}
\usepackage[noend]{algpseudocode}
\usepackage{bigints}
\usepackage[mathscr]{euscript}
\usepackage[makeroom]{cancel}
\usepackage{graphicx}
\usepackage[table]{xcolor}
\usepackage{subcaption}
\usepackage{enumitem}
\usepackage{verbatim}
\usepackage[font={small,it,singlespacing}]{caption}
\usepackage{lscape}
\usepackage{caption}
\usepackage{hyperref}
\usepackage[sort]{natbib}
\usepackage{parskip}
\usepackage{array}
\usepackage{multirow}
\usepackage{graphicx}
\usepackage{wrapfig}
\usepackage{lscape}
\usepackage{rotating}
\usepackage{epstopdf}
\usepackage[compact]{titlesec}
\usepackage{makecell}
\usepackage{authblk}
\usepackage[toc,page]{appendix}
\usepackage{setspace}
\usepackage{xr}

\usepackage{color, colortbl}
\definecolor{Gray}{gray}{0.9}

\usepackage{algorithm}% http://ctan.org/pkg/algorithm
\usepackage{algpseudocode}% http://ctan.org/pkg/algorithmicx

\usepackage{boldline}

\hypersetup{colorlinks,citecolor=blue}
\graphicspath{{./images//}}

\usepackage[sort]{natbib}

\setlist{nosep}

\sectionfont{\fontsize{13}{13}\selectfont}
\subsectionfont{\fontsize{12}{12}\selectfont}

\newcolumntype{L}[1]{>{\raggedright\let\newline\\\arraybackslash\hspace{0pt}}m{#1}}
\newcolumntype{C}[1]{>{\centering\let\newline\\\arraybackslash\hspace{0pt}}m{#1}}
\newcolumntype{R}[1]{>{\raggedleft\let\newline\\\arraybackslash\hspace{0pt}}m{#1}}

\theoremstyle{plain}

\newtheoremstyle{exampstyle}
  {\topsep} % Space above
  {\topsep} % Space below
  {} % Body font
  {} % Indent amount
  {\bfseries} % Theorem head font
  {} % Punctuation after theorem head
  {.5em} % Space after theorem head
  {} % Theorem head spec (can be left empty, meaning `normal')

\newtheoremstyle{exampstyle}
  {\topsep} % Space above
  {\topsep} % Space below
  {} % Body font
  {} % Indent amount
  {\bfseries} % Theorem head font
  {.} % Punctuation after theorem head
  {.5em} % Space after theorem head
  {} % Theorem head spec (can be left empty, meaning `normal')

\usepackage{algpseudocode}
\usepackage{algorithm}

\algnewcommand\algorithmicinput{\textbf{Input:}}
\algnewcommand\Input{\item[\algorithmicinput]}

\algnewcommand{\algorithmicoutput}{\textbf{return:}}
\algnewcommand\Output{\item[\algorithmicoutput]}

\algnewcommand\algorithmicforeach{\textbf{for each}}
\algdef{S}[FOR]{ForEach}[1]{\algorithmicforeach\ #1\ \algorithmicdo}

\theoremstyle{plain}

\title{\normalsize{\textbf{How Should We Measure Empirical Risk when Synthesizing Population Data?}}\vspace{-1em}}

\author{\small{Joshua Snoke}}

\affil{\textit{Massive Data Institute, Georgetown University} \\ \textit{joshua.snoke@georgetown.edu}}
\date{}

\begin{document}
\doublespacing
\maketitle

\vspace{-30pt}

\noindent\textbf{Abstract:} Synthetic data has become a prominent solution for preserving privacy while sharing data, but current empirical risk assessment frameworks fundamentally assume a sample-based context that fails to translate for the evaluation of synthetic population level datasets. This commentary explores the implications when synthesizing entire populations in order to do population-level data science, arguing that traditional metrics, such as Membership Inference Attacks (MIA) and Attribute Inference Attacks (AIA), require re-examination. First, MIA may be rendered irrelevant in contexts where population membership is public knowledge or not considered sensitive information. Second, the risk of singling out is heightened because the confidential data contain full population information. Additionally, the absence of an ``out-of-sample" comparison group for attribute inference means we need to define other policies when defining acceptable inferences. Finally, we cannot rely on simply returning to subsampling prior to generating synthetic data if the use case is truly to enable population-level data science. This commentary highlights the necessity for considering context when generating and evaluating synthetic population data.

\section{Introduction}\label{introduction}

Synthetic data has emerged as one of the most popular methods for releasing data while preserving privacy and confidentiality. From its early beginnings, it was proposed as an adaptation of multiple imputation for missing data to help solve the issue of creating public-use microdata in the context of official statistics \citep{rubin1993statistical}. Synthetic data has since expanded to a wide range of applications, from synthetic surveys and administrative records \citep{abowd2006final} to synthetic electronic health records \citep{zhang2021synteg} or images \citep{chung2025sok} to name only a couple examples.

In spite of this popularity, synthetic data remains polarizing for many individuals, particularly when it comes to measuring the disclosure risk of releasing synthetic data \citep{desfontaines2024empirical,stadler2022synthetic,guepin2023synthetic}. The question of how to measure risk from the release of synthetic data has recently gained greater attention, but unless one takes the approach of guaranteeing privacy as part of the process (e.g., through the framework of differential privacy\footnote{I do not address differentially private synthetic data directly in this article, since it is a fundamentally different privacy framework. While it still makes sense sometimes to apply empirical measures to describe the protections offered by a differentially private data set, it is not essential for defining privacy loss.}), there is still little consensus on how to empirically measure the risk of synthetic data.

Part of the issue stems from the fact that synthetic data has been applied to such a wide range of data contexts in a short period of time. While the simple concept of generating ``synthetic'' records from a model that captures aspects of the confidential data can be easily extended across domains, the concept of measuring risk is not so straightforward. Off-the-shelf tools for evaluating synthetic data are often applied without taking special care to consider whether the metrics applied make sense for the context from which the confidential data was generated.

One such context is population data, which poses a unique departure from the original (and prevailing) concept of synthetic data. In this commentary, I assume a scenario where the goal is to take a population-level data set and create a synthetic replacement that enables the same type of population-level data science while preserving privacy for those individuals in the data. I focus on the question of empirically measuring risk in this context. Measuring utility is important and similarly faces questions about its adaptability to the population context, but given the significant imbalance in the attention given to measuring utility over risk in synthetic data, I have chosen to focus on risk.

Synthetic data methods fundamentally assume the confidential data represent a sample, both in the synthetic data generators but \emph{also} in the methods for measuring empirical risk. This requires us to use caution when applying the normal approaches; we need to examine the underlying assumptions and only apply the methods in ways that are appropriate for population data. Ultimately, this is a context that the field has not sufficiently addressed and more work will be needed to appropriately measure risk.

\section{Brief History of Synthetic Data}\label{brief-history-of-synthetic-data}

It is helpful to provide a short review of the history of synthetic data, since the current methodology is built upon fundamental assumptions that go back to the first proposals. In this section, I draw upon an excellent review article published recently \citep{drechsler2024years}, and I strongly encourage readers to read the whole piece.

As was already referenced, synthetic data owes its earliest lineage to work in official statistics that drew on concepts from multiple imputation for missing data \citep{rubin1993statistical,little1993statistical}. Rubin's original idea was to use the confidential survey records to develop a model from which a synthetic population could be simulated and then multiple synthetic data sets could be sampled and released. Eventually the last two steps were combined. Over time different variants were proposed which depended on the assumed risk and goals of the synthetic data, including partially synthetic data \citep{reiter2003inference} (i.e., only synthesizing certain parts of the population) and only creating a single synthetic data set \citep{raab2016practical}.

As synthetic data entered its second and third decades, the research field focused primarily on methods for ensuring high analytic quality of the synthetic data \citep{reiter2005using,drechsler2011empirical,snoke2018general}. While there were proposals for measuring risk \citep{reiter2014bayesian}, empirical evaluations were rarely conducted and practitioners largely relied on the idea that because the records were ``fake'', they did not present a risk\footnote{This was especially true for fully synthetic data. More risk measures were proposed for partially synthetic data, but these are largely not applicable in the current paradigm where most applications involve synthesizing all records.}. In this period, synthetic data also saw its first institutional applications, still focused in the realm of statistical agencies and official statistics \citep{abowd2006final,benedetto2018creation,kinney2011towards,kinney2014synlbd,drechsler2012new,nowok2017providing}.

Roughly ten years ago synthetic data experienced a substantial shift, due to the increased applicability of deep learning models. Following the proposal of Generative Adversarial Networks (GANs) and Variational Autoencoders (VAEs), researchers and practitioners across academia, industry, and government realized that synthetic data could be applied to much more high-dimensional and complex data than had previously been considered \citep{camino2018generating,ma2020vaem}. Now with the advent of even more capable AI models, this trend has only accelerated \citep{jordon2022synthetic,waseem2026review}. Synthetic data also expanded beyond confidentiality into data augmentation, primarily in settings where companies build large machine learning models which require more training samples than have been collected.

The rise of deep learning synthesis models helped the community perceive the risk inherent in synthetic data. While the records are ``fake'', a model which memorizes and replicates records from the confidential data will not preserve privacy. With parametric models, or the less complex earlier machine learning models, it seemed unlikely that replicating records was a significant issue, but with increasingly large models that are more prone to overfit the data, this risk became a reality. This led to a recognition in the past decade of the need to empirically measure the disclosure risk from releasing synthetic data.

\section{Defining Risk for Synthetic Data and the Assumption of Sampling}\label{defining-risk-for-synthetic-data-and-the-assumption-of-sampling}

Before diving into specific challenges arising with synthesizing population data, it is helpful to review the general approaches to defining and measuring risk with synthetic data. Classically, risk has been divided into \textbf{identification} and \textbf{attribute} risk \citep{pilgram2025consensus,kaabachi2025scoping}. These are broad categories but they encompass commonly understood harms. For fully synthetic data, traditional re-identification which links a particular data record to an individual is not considered meaningful. Instead, identification risk is normally associated with the potential harm from learning that an individual contributed to the confidential data. This differs from traditional re-identification, but it can be harmful if participation in the data is linked to a sensitive reason e.g., a survey of HIV positive individuals. Attribute risk is connected to the harm of learning previously unknown information about an individual. Both risks are typically considered with respect to what could have been learned without the synthetic data, sometimes called the baseline information.

Identification risk is commonly measured using a Membership Inference Attack (MIA), which measures an attacker's ability to determine if a specific record was in the training data \citep{stadler2022synthetic}. This is typically measured by using the distance between targeted and synthetic records to predict matches \citep{elemam2022validating} or by creating prediction models for the existence of input records based on replicating the synthesis process \citep{stadler2022synthetic}. Attribute inference is measured by going one step further, conditioning on the prediction that an individual with certain record values participated in the data, and trying to estimate additional (sensitive) attributes for those partial records. This is normally done either by predicting attribute values based on ``matched'' records or by using prediction models based on partial records.

Though not often applied to synthetic data, the concept of \textbf{singling out} also merits mentioning. This is the ability to isolate a record based on unique characteristics---a key metric within legal frameworks like the GDPR \citep{eu2016gdpr,cohen2020towards}. We may be interested in measuring this risk for synthetic data as an extension of the harms from membership disclosures. If it is possible to infer, with high accuracy, that an individual participated in the confidential data and also has a unique set of characteristics, that might meet the definition of singling out and constitute a relevant disclosure. This consideration is amplified in the context of population data.

Membership inference and attribute inference, as typically operationalized, assume that the synthetic data are generated based on a sample from a broader population. These assumptions are often implicit, but they have implications if we want to evaluate the risk for synthetic population data. We need to dig further into these issues and be careful to avoid a few mistakes before applying these techniques.

\section{Avoiding Pitfalls When Applying MIA on Population Data}\label{avoiding-pitfalls-when-applying-mia-on-population-data}

When we apply MIA to evaluate identification risk for synthetic population data, we must ask if inferring membership represents a meaningful disclosure at all. This depends on two related questions: (1) is membership in a population considered public knowledge and (2) is membership in the population considered sensitive? Both questions must be answered in the affirmative for membership inference to constitute a meaningful risk. The second question is one that must be asked for samples as well, but it is less common for an entire population to be considered sensitive.

In some settings, knowing someone is in a database, such as participation in a sensitive biomedical study, is a clear privacy breach \citep{zhang2022membership}. But generally speaking this is not what is meant by population data. Typically sensitive groups are subsets of broader populations, such as identifying an abortion cohort within population-level administrative health data \citep{schummers2022accurate}. Additionally, for many populations, such as the U.S. Decennial Census, participation is assumed to be public knowledge. Generally speaking, we can expect it is rare to affirm both questions in most applications of population data science.

If membership is already a matter of public record, MIA becomes irrelevant. We still may want to evaluate the risk of inferring individuals within sensitive groups in the data, but it is no longer accurate to call this membership inference. More often than not, the true risk in a population context lies not in membership, but in attribute inference.

In situations where membership does constitute a meaningful disclosure, a second issue arises from the commonly used methods for estimating membership disclosure risk. Specifically, certain methods partition the data as part of membership attacks \citep{elemam2022validating} in order to accurately assess the ability of an attacker who does not possess information about the entire population. Unfortunately, this assumes the confidential data come from a sample, so this approach no longer works for population data! Other methods exist which do not partition the data \citep{guepin2023synthetic}, but caution should be exercised when applying MIA to ensure that the underlying assumptions match the context.

\section{Considering the Heightened Risk of Singling Out}\label{considering-the-heightened-risk-of-singling-out}

When inferring membership does not constitute a disclosure, there may still be risk from (inferred) singling out of records. In fact, this risk is heightened with synthetic population data due to the simple fact that the risk of singling out is heightened with population data. In the context of samples, singling out a record requires determining uniqueness in the population conditional on uniqueness in the sample. Though this can be done \citep{rocher2019estimating,skinner2008assessing}, it adds a level of protection.

Conversely, with population data singling out requires only determining unique characteristics in the data at hand. Population-level synthetic data that captures features of the confidential data provides direct insights into how unique a record is within the population. Accordingly, creators of population synthetic data should be careful when considering which quasi-identifiers are included and whether they allow for singling out of specific individuals. This risk is not captured in the dominant evaluation metrics (i.e., membership inference, attribute inference), and it represents a current weakness in the synthetic risk evaluation toolbox. More work should explicitly consider how to measure the risk of singling out with synthetic data, particularly when using population data which heightens the risk from this type of attack.

\section{Making Policy Decisions about Acceptable Attribute Inferences}\label{making-policy-decisions-about-acceptable-attribute-inferences}

Perhaps the most significant challenge when it comes to assessing real risks for synthetic population data is distinguishing between attribute inferences that are due to releasing the synthetic data versus what could be inferred without it. A commonly understood issue in measure attribute inference risk for synthetic data is differentiating between inferences that simply reveal information about the population versus revealing information about individuals in the data \citep{pilgram2025consensus}. One approach for disentangling these two is to compare the accuracy of inferences for individuals who are in the data versus those who are not, similar to the hold-out data approach for membership inference described earlier. If we do not observe a meaningful increase in the accuracy of inferences for individuals in the data used to create synthetic data, then the argument goes that the synthetic data do not increase attribute disclosure risk.

Unfortunately, this argument falls apart in the face of population-level data. We can no longer hide separate inferences for those ``in the data'' or ``not in the data'' if we are truly working with population-level data\footnote{Even if we consider the fact that population-level data rarely actually includes every single individual in the population, this issue remains as long as we are working with a near census.}. In a population, \emph{there is no one outside the data}. This reality forces us instead to need to make policy decisions; we must explicitly decide which inferences are acceptable for the goals of the study and which constitute a disclosure. For example, perhaps acceptable inferences depend on the size of the group about which we are learning, tying it back to the concept of singling out risk. Alternatively, the policy decision may revolve around the harms that could come as a result of the disclosures or the potential vulnerability of the individuals. Or we may choose to define acceptable disclosures based on the heterogeneity of the outcome, which goes back to the idea of constructing a baseline (prior) risk. These are simply ideas, not recommendations. This type of assessment is not part of the normal synthetic data generation process, so the research field and policymakers will need to develop more standard guidance for making these decisions.

\section{Is Subsampling the Solution?}\label{is-subsampling-the-solution}

Finally, one might read this and think that the solution is simply to sample from the population \emph{before} creating synthetic data. This idea may suffice and allow us to apply the usual risk metrics as if the confidential data came from a sample. But in order to do this, we need to determine the purpose of the synthetic data and to be clear about our assumptions in the generation process. Specifically, will the synthetic data be used for population-level data science? Is the goal of the synthetic data to act as a replacement for the confidential population data, or is it simply to provide a sample data set? If it is the former, and we use a sample to generate the synthetic data, we need to be explicit that we are arguing that it is possible to create a useful synthetic population which is based on a sample of our confidential data. If not, it is likely not appropriate for the use case to create the synthetic data based on a sample.

\section{Conclusion}\label{conclusion}

In many ways, this commentary boils down to recognizing that context matters when applying evaluation metrics for synthetic data. We must recognize that risk measures designed for samples will not always translate to population data. In particular, we need to be aware of the implicit assumptions underlying membership inference and determine whether this inference constitutes a disclosure for population data. We also need to consider the increased risk from singling out and take appropriate steps to measure this type of risk. Along with these, we must take explicit policy choices about which attribute inferences are acceptable and which ones constitute disclosures in our context. Finally, we should be careful if subsampling to ensure that the appropriate uses of the synthetic data are clear. By paying attention to these issues, we can increase the likelihood that our synthetic data evaluation is meaningful.

\section{Disclosure Statements}\label{disclosure-statements}

The author declares there are no conflicts of interest.

The commentary did not require review by an ethics committee because no
human subjects were involved.

No data are available as part of this commentary.

The author used Gemini 3 through Georgetown University's license for
editing, summarization, and citation formatting purposes. The author
originally wrote and reviewed everything in the commentary for accuracy.

\bibliographystyle{chicago}
\bibliography{ref}

\end{document}